\documentclass[reprint,twocolumn,showpacs,superscriptaddress]{revtex4-1}
\usepackage[dvips]{graphicx}
\usepackage{mathrsfs}
\usepackage{amsmath}
\usepackage{amssymb}
\usepackage{amsfonts}
\usepackage{ae}
\usepackage{units}

\begin{document}
%
\title{Ground-state cooling of a suspended nanowire through inelastic macroscopic quantum tunneling in a current-biased Josephson junction}
\author{Gustav Sonne}
\email{gustav.sonne@physics.gu.se}
\affiliation{University of Gothenburg, Department of Physics, SE-412 96 G\"oteborg, Sweden}
\author{Leonid Y. Gorelik}
\affiliation{Chalmers University of Technology, Department of Applied Physics, SE-412 96 G\"oteborg, Sweden}
\date{\today}
%
\begin{abstract}
We demonstrate that a suspended nanowire forming a weak link between
two superconductors can be cooled to its motional ground state by a
supercurrent flow. The predicted cooling mechanism has its origins in
magnetic field induced inelastic tunneling of the macroscopic
superconducting phase associated with the junction. Furthermore, we
show the voltage-drop over the junction is proportional to the average
population of the vibrational modes in the stationary regime, a
phenomena which can be used to probe the level of cooling.
\end{abstract}
%
\maketitle

Nanoelectromechanical systems (NEMS) are fast approaching the limits
set by quantum
mechanics~\cite{Schwab2005,Blencowenano,Blencowe2004}. Achieving such
conditions requires that the mechanical subsystem can be brought into,
and detected, in its quantum mechanical ground state.  In general this
condition demands that an energy quanta associated with the mechanical
motion is much larger than the energy associated with the thermal
environment. For an oscillator with a mechanical frequency of
\unit[100]{MHz} this implies temperatures as low as a few
\unit[]{mK}. However, using oscillators with higher mechanical
frequencies the quantum limit can be reached, as recently demonstrated
by O'Connell {\it et al.}~\cite{OConnell2010}.

The most common device geometries of NEMS to date consist of
mechanical oscillators in the form of cantilevers, suspended beams or
microtoroids.  These typically have much lower resonance frequencies
than those reported in Ref.~\cite{OConnell2010}, hence reaching the
quantum limit in these devices is very challenging. To circumvent this
problem, back-action cooling of the mechanically compliant element is
often employed whereby the number of mechanical vibrons is reduced
without necessarily lowering the ambient temperature.  Suggestions for
different cooling mechanisms are plentiful, see
e.g. Refs.~\cite{Martin2004,Ouyang2009,WilsonRae2004,Zippilli2009,Sonne2010}. Common
to these is that the oscillator is cooled either by coupling its
mechanical oscillations to electromagnetic photons or a flow of charge
carriers.

In the present paper we suggest a new mechanism of cooling not
previously considered and show that ground-state cooling of the
mechanical oscillator is possible. Considering the nanomechanical
oscillator as a weak link in a current-biased Josephson junction we
show that we can access a regime analogous to the resolved side-band
limit~\cite{Schliesser2008}, whereby the number of mechanical vibrons
in the system can be reduced by a factor of $\sim$\unit[100]{}. In the
limit of a high mechanical quality factor the resulting vibron
population is shown to be well within the quantum regime.
\begin{figure}
\includegraphics[width=0.45\textwidth]{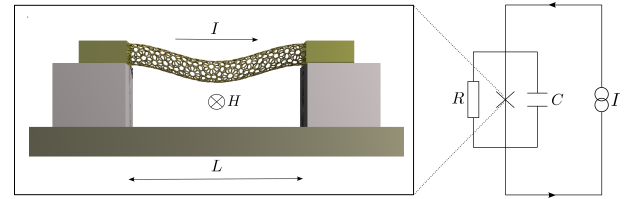}
\caption{(Color online) Schematic diagram of the system. (Left) A
  suspended nanowire of length $L$ forms a weak link between two
  current-biased superconducting leads. The transverse magnetic field
  $H$ is applied perpendicular to the nanowire. (Right) The equivalent
  electronic circuit. A constant current $I$ is applied to the
  Josephson junction which is connected in parallel to a capacitor $C$
  and a resistance $R$.}
\label{picture}
\end{figure}

The cooling mechanism considered here is achieved by coupling the
mechanical vibrations of the oscillator to the supercurrent through
the junction. Below we show that the suggested setup not only allows
for ground-state cooling of the mechanical oscillator, but
simultaneously probes the macroscopic nature of the superconducting
phase associated with the junction. As such, the proposed system
allows for interesting physical observations on both the mechanical
and the electronic subsystems.

Figure~\ref{picture} shows a schematic picture of the system
considered. It consists of a metallic carbon nanotube suspended over
two superconducting leads biased at a current $I$. Transverse to the
in-plane motion of the nanotube a magnetic field $H$ is applied which
induces coupling between the bending modes of the wire to the
supercurrent through it \cite{drivenoscillator}. Below we analyze the
influence of the electromechanical coupling and show that for resonant
current-biased conditions this may lead to ground-state cooling of the
vibrations of the nanowire.

In our analysis we restrict the description of the mechanical degrees
of freedom of the nanowire to the fundamental bending mode, which is
considered as a harmonic oscillator with frequency $\omega$. The
Hamiltonian describing the system presented in Fig.~\ref{picture} has
the form,
\begin{gather}
\label{finalH}
\!\!\!\!\hat{\mathcal{H}}= 4E_c\hat{n}^2-j\hbar\hat{\phi}-E_J\cos(\hat{\phi}-\Phi\hat{u})+\hbar\omega\hat{b}^{\dagger}\hat{b}\,.
\end{gather}
Here, $\hat{n}$ is the operator for the number of Cooper pairs on the
junction and $\hat{\phi}$ is the corresponding operator for the
superconducting phase \cite{IngoldNazarov}
($[\hat{\phi},\hat{n}]=i$). In \eqref{finalH}, $E_c=e^2/(2C)$ is the
Coulomb energy where $C$ is the capacitance of the junction,
$j=I/(2e)$ is the flow of Cooper pairs and $E_J$ is the Josephson
energy. The operators $\hat{b}^{\dagger}$ $[\hat{b}]$ are creation
[annihilation] operators for the oscillator where
$\hat{u}=\hat{b}+\hat{b}^{\dagger}$ is the dimensionless deflection of
the wire. In the above, the parameter $\Phi=4g\pi LHu_{zp}/\Phi_0$
characterizes the strength of coupling between the mechanical and
electronic degrees of freedom. Here, $u_{zp}=(\hbar/(2m\omega))^{1/2}$
is the zero-point amplitude of the nanowire, $m$ and $L$ is the
effective mass and length of the suspended part of the wire
respectively, $\Phi_0=\pi\hbar/e$ is the flux quantum and $g$ is a
numerical factor of the order of unity which accounts for the profile
of the fundamental bending mode \cite{Shekhter}.

The third term in \eqref{finalH} describes on the one hand the Lorentz
force on the nanowire induced by the Josephson current. On the other
hand, it gives the deflection-dependence of the Josephson current due
to the motion of the wire in the magnetic field
\cite{drivenoscillator}. In what follows we consider a nanotube of
length $L\sim 1$\unit[]{$\mu$m}, for which $u_{zp}\lesssim 1$\unit[]{Å},
in a magnetic field $H\sim 1$\unit[]{T}. With these parameters
$\Phi\lesssim 0.3$, and we consider only the linear terms in the
expansion of \eqref{finalH} with respect to $\Phi$. With this
expansion the Hamiltonian reads,
\begin{gather}
\label{hamil}
\hat{\mathcal{H}}=\hat{\mathcal{H}}_J+\hat{\mathcal{H}}_{m}+\hat{\mathcal{H}}_{int}\,,\\
\hat{\mathcal{H}}_J=4E_c\hat{n}^2-j\hbar\hat{\phi}-E_J\cos\hat{\phi}\,, \qquad \hat{\mathcal{H}}_{m}=\hbar\omega\hat{b}^{\dagger}\hat{b}\,,\notag\\
\hat{\mathcal{H}}_{int}=-E_J\Phi(\hat{b}^{\dagger}+\hat{b})\sin\hat{\phi}\notag\,.
\end{gather}
Here, $\hat{\mathcal{H}}_J$ is the Josephson Hamiltonian, which under
the condition $j<E_J/\hbar$, describes the electronic subsystem in the
so-called tilted washboard potential. In \eqref{hamil},
$\hat{\mathcal{H}}_{int}$ describes the interaction between the
mechanical and electronic subsystems with $\hat{\mathcal{H}}_m$ the
Hamiltonian of the former.

\begin{figure}
\includegraphics[width=0.35\textwidth]{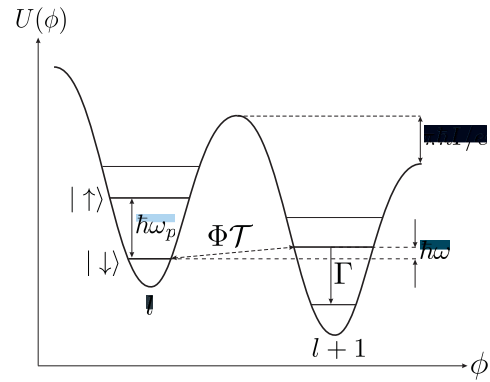}
\caption{(Color online) Schematic diagram of the tilted washboard
  potential $U(\phi)=-E_J\cos\phi-j\hbar\phi$ as a function of phase
  $\phi$ at current-bias $I=e/\pi(\omega_p-\omega)$. Here, $l$ labels
  the valleys of the potential and $\sigma=\uparrow,\downarrow$ are
  the two energy levels within the valleys considered. In the above,
  $\Phi\mathcal{T}$ is the inelastic tunneling amplitude between two
  energy levels in consecutive valleys.  The quantity $\Gamma$ is the
  transition rate from the second to the first level within a valley
  generated by interactions with the quasiparticle environment (see
  text).}
  \label{tilted}
\end{figure}

Below we will take the Columb energy to be much smaller than the
Josephson energy, $4E_c/E_J\ll 1$. This condition implies that the
characteristic interlevel distance between the quantized states of the
Josephson junction associated with a given local minimum of the
washboard potential, $\hbar\omega_p=(8E_JE_c)^{1/2}$, is much smaller
than the height of the barrier separating different local
minima. Here, $\omega_p$ is the plasma frequency of the junction. We
also take the external temperature $T$ to be low,
$T<\hbar\omega_p/k_B$, such that transitions between states associated
with different local minima can only occur through under-barrier
tunneling. A schematic diagram of the quantum state of the electronic
subsystem described through $\hat{\mathcal{H}}_J$ is shown in
Fig.~\ref{tilted}.

Under-barrier tunneling between two consecutive valleys in
Fig.~\ref{tilted} changes the state of the Josephson junction through
the associated change of the phase. Such tunneling events, commonly
referred to as macroscopic quantum tunneling (MQT), are greatly
enhanced if the two energy levels involved in the transition are in
resonance. This can be achieved by tuning the current-bias. Thus, we
define the critical bias current $I^*$ as the current which ensures
that the lowest (first) level in a given valley is resonant with the
second level in the next valley, $I^*\simeq e\omega_p/\pi$
\cite{Schmidt1991}.  As the potential defined by $\hat{\mathcal{H}}_J$
is only to first approximation parabolic, the spacing between the
energy levels within a given valley is not constant. As such, we will
in the following only consider tunneling between the two lowest
electronic states and neglect any coupling to higher levels. This is
justified as the, e.g., the second and third levels are far from
resonance if the junction is biased at $I\simeq I^*$ (see
Fig.~\ref{tilted}) \cite{Schmidt1991}.

The electronic system in Fig.~\ref{tilted} is coupled to the
mechanical subsystem by the magnetic field. As such, MQT can in the
present situation also be accompanied with the emission/absorption of
a quanta of mechanical energy, $\hbar\omega$. Performing a WKB
analysis for the MQT amplitude we find that the overlap integrals for
the inelastic channels is of the order of $\Phi\mathcal{T}$ where
$\mathcal{T}\propto\hbar\omega_p\exp(-\pi(E_J/(2E_c)^{1/2})<\hbar\omega$
is the tunneling amplitude in the elastic channel. Here, we note that
the $\phi$-dependence of $\hat{\mathcal{H}}_{int}$ only leads to a
renormalization of the parameter $g$ in the definition of $\Phi$. Also
note that due to the large separation in energy, $\omega\ll\omega_p$,
the electromechanical coupling will not introduce additional tunneling
channels between the higher electronic energy levels.

The inelastic tunneling channels change the number of mechanical
vibrons such that cooling of the oscillator is possible if transitions
through the absorption channel can be promoted. Below we show that
this can be achieved by tuning the bias current so that the absorption
channel is resonant; the first level in a valley $l$ is separated by
$\hbar\omega$ from the second level in $l+1$ as shown in
Fig.~\ref{tilted}. A further condition for cooling is that the
electronic subsystem, once in the second energy level, relaxes to the
lower level at a rate $\Gamma$ which is faster than the rate at which
the system tunnels back with the emission of a vibron,
$\Gamma>\mathcal{T}/\hbar$. Such relaxation arises due to interaction
with the quasiparticle environment as discussed further below.

To perform a quantitative analysis of the system we introduce the
basis $\vert l,\sigma\rangle$ where $l$ labels the valleys of the
potential and $\sigma=\uparrow,\downarrow$ labels the energy levels
inside a given valley ($\downarrow$ is the first and $\uparrow$ is the
second level). In this basis the Hamiltonian reads,
\begin{gather}
\label{H}
\hat{\mathcal{H}}=\hat{\mathcal{H}}_0+\hat{\mathcal{H}}_{\mathcal{T}}\,,\\
\hat{\mathcal{H}}_0=\hat{\mathcal{H}}_J+\hat{\mathcal{H}}_m=\sum_{l,n,\sigma}(\mathcal{F}_{l,\sigma}+\hbar\omega\hat{b}^{\dagger}\hat{b})\vert l,\sigma\rangle\langle l,\sigma\vert \,,\notag\\
\hat{\mathcal{H}}_\mathcal{T}=\sum_{l}\mathcal{T}\left(\Phi(\hat{b}+\hat{b}^{\dagger})+1\right)\vert l+1,\uparrow\rangle\langle l,\downarrow\vert+\textrm{h.c.}\,.\notag
\end{gather}
In the above, $\mathcal{F}_{l,\sigma}=\hbar\omega_pm_\sigma-l\pi\hbar
I/e$ are the eigenvalues for the electronic degrees of freedom in the
basis $\vert l,\sigma\rangle$, where $m_\uparrow=1$ and
$m_\downarrow=0$. From the form of the Hamiltonian \eqref{H} one can
see that due to the electromechanical coupling the number of vibrons
in the system is not conserved and may change due to macroscopic
tunneling of the electronic system from one valley to the next.

To describe the joint dynamics of the electronic and mechanical
degrees of freedom we will start our analysis from the Liouville-von
Neumann equation for the density matrix $\hat{\rho}$ of the system,
\begin{align}
\label{vonNeum}
\frac{\partial \hat{\rho}}{\partial t}=-\frac{i}{\hbar}&\left[\hat{\mathcal{H}}_0+\hat{\mathcal{H}}_{\mathcal{T}},\hat{\rho}\right]+\hat{J}(\hat{\rho})+\notag\\
&\gamma(1+n_B)\mathcal{L}_{\hat{b}}(\hat{\rho})+\gamma n_B\mathcal{L}_{\hat{b}^{\dagger}}(\hat{\rho})\,.
\end{align}
Here, $\hat{J}(\hat{\rho})$ is a phenomenological damping operator for
the electronic system \cite{Schmidt1991},
\begin{align}
\hat{J}(\hat{\rho})=-&\frac{\Gamma}{2}\left(\sum_{l}\vert l,\uparrow\rangle\langle l,\uparrow\vert\hat{\rho}+\hat{\rho}\vert l,\uparrow\rangle\langle l,\uparrow\vert\right)+\notag\\
\Gamma&\sum_{l,l'}\vert l,\downarrow\rangle \langle l,\uparrow\vert \hat{\rho}\vert l',\uparrow\rangle\langle l',\downarrow\vert\,.
\label{damping}
\end{align}
In the equivalent circuit scheme (see Fig.~\ref{picture}) this damping
derives from the parallel resistance $R$, which in the present
situation causes the system to decay from the $\uparrow$ state to the
$\downarrow$ state in a given valley.  In \eqref{damping},
$\Gamma=\omega_p/Q_{el}$ is the electronic damping rate, where
$Q_{el}=\omega_pRC$ is the corresponding quality factor. Here we
consider $Q_{el}\gg~1$ which implies that the influence from the
electronic quasiparticle environment on the tunneling processes is
negligible~\cite{Schmidt1991,Esteve1986,Hatakenaka1990}. We will
further suppose that the quality factor $Q_{el}$ is so large that
broadening of the second energy level,
$\Delta\omega_p=\omega_p/(2Q_{el})$, is small enough for the inelastic
resonance transitions to be resolved, $\Delta\omega_p<\omega$.

The second damping term in \eqref{vonNeum},
$\mathcal{L}_{\hat{a}}(\hat{\rho})=(2\hat{a}\hat{\rho}\hat{a}^{\dagger}-\hat{a}^{\dagger}\hat{a}\hat{\rho}-\hat{\rho}\hat{a}^{\dagger}\hat{a})/2$,
is the standard Lindblad operator which models interactions between
the oscillator and the thermal environment. Here, $\gamma=\omega/Q$ is
the mechanical damping rate with $Q$ the quality factor and
$n_B=(\exp(\beta\hbar\omega)-1)^{-1}$, where $\beta=(k_BT)^{-1}$, is
the average number of vibrons in thermal equilibrium.

Below we investigate the stationary solution to \eqref{vonNeum}. To
find this solution we perform a standard perturbative analysis in the
small parameters
$\mathcal{T}/(\hbar\Gamma),\gamma/\Gamma\propto\epsilon\ll 1$ and look
for a solution of the density matrix of the form
$\hat{\rho}=\hat{\rho}_0+\epsilon\hat{\rho}_1+\epsilon^2\hat{\rho}_2...$
(for a full derivation of the results presented below
see Appendix~\ref{append}). Substituting this into
\eqref{vonNeum} one finds that the leading order solution
$\hat{\rho}_0$ has the form $\hat{\rho}_0=\sum_{l,n}\vert
l,\downarrow,n\rangle \rho_0(l,\downarrow,n)\langle
l,\downarrow,n\vert$, where the index $n$ labels the Fock state of the
oscillator. From \eqref{vonNeum} we also find the first order
correction $\hat{\rho}_1=\sum_{l,n,j=-1,0,1}\vert
l+1,\uparrow,n+j\rangle c_j(l,n)\langle l,\downarrow,n\vert
+\textrm{h.c.}$ where the sum $\sum_lc_j(l,n)\equiv C_j(n)$ satisfy
the following relation,
\begin{gather}
C_j(n)=\frac{\mathcal{T}_j^{(n)}P(n)}{-\Delta\mathcal{F}_j+i\hbar\Gamma/2}\,,\qquad
\!\!\!\!\!\!\!\!\!\mathcal{T}_j^{(n)}=\left\{\begin{array}{l}
\mathcal{T}_+^{(n)}=\mathcal{T}\Phi\sqrt{n+1}\\\mathcal{T}_0^{(n)}=\mathcal{T}\\\mathcal{T}_-^{(n)}=\mathcal{T}\Phi\sqrt{n}\,.\end{array}\right.\notag
\end{gather}
Here, $P(n)=\sum_l\rho_0(l,\downarrow,n)$ is the population of the
vibrational modes of the oscillator. Developing the perturbative
expansion one finds that the equation for the second order term,
$\hat{\rho}_2$, can only be resolved if $P(n)$ satisfy the following
equation,
\begin{align}
(\Gamma_-+&\gamma(1+n_B))[(n+1)P(n+1)-nP(n)]+\notag\\
&(\Gamma_++\gamma n_B)[nP(n-1)-(n+1)P(n)]=0\,.
\label{rho0}
\end{align}
Here, $\Gamma_j$ are the different tunneling rates; $j=-,0,+$ are
respectively the absorption, elastic and emission channel,
\begin{gather}
\label{rates}
\Gamma_{\pm}=\Gamma\frac{4\Phi^2\mathcal{T}^2}{4(\Delta\mathcal{F}_\pm)^2+\hbar^2\Gamma^2}\,,\qquad \Gamma_{0}=\Gamma\frac{4\mathcal{T}^2}{4(\Delta\mathcal{F}_0)^2+\hbar^2\Gamma^2}\,,\notag\\
\Delta\mathcal{F}_{0}=\mathcal{F}_{l+1,\uparrow}-\mathcal{F}_{l,\downarrow}\,,\qquad
\Delta\mathcal{F}_{\pm}=  \mathcal{F}_{0}\pm\hbar\omega\,.\notag
\end{gather}

Considering the operator for the potential over the Josephson junction
$\hat{V}=i[\hat{\mathcal{H}},\hat{\phi}]/(2e)$ (in our representation
$\hat{\phi}=2\pi\sum_{l,\sigma}\vert l,\sigma\rangle l\langle
l,\sigma\vert$) we find,
\begin{equation}
\label{volt}
\hat{V}=\frac{\pi}{ie}\sum_l\mathcal{T}\left(\Phi(\hat{b}+\hat{b}^{\dagger})+1\right)\vert l+1,\uparrow\rangle\langle l,\downarrow\vert+\textrm{h.c.}\,.
\end{equation}
This implies that the stationary bias voltage,
$V=\textrm{Tr}(\hat{V}\hat{\rho})$, is zero to leading order in
$\hat{\rho}$. Thus, the potential drop is given by the first order
correction to the density matrix,
$V=\textrm{Tr}(\hat{V}\hat{\rho}_1)$, which implicitly depends on the
coefficients $C_j(n)$. Solving equation \eqref{rho0} we find that the
average number of vibrons, $\langle n\rangle=\sum_n nP(n)$, is given
by,
\begin{gather}
\label{nav}
\langle n\rangle =\frac{n_B\gamma+\Gamma_+}{\gamma+\Gamma_--\Gamma_+}\,,
\end{gather}
and that the voltage drop scales with $\langle n\rangle$ as,
\begin{gather}
\label{volt2}
V=\frac{\pi\hbar}{e}\left(\Gamma_-\langle n\rangle+\Gamma_0+\Gamma_{+}(\langle n\rangle +1)\right)\,.
\end{gather}

Here we note that the potential drop in the stationary regime is
primarily determined by the elastic tunneling rate, $\Gamma_0$. This
is consistent with the physical processes discussed, i.e. in the limit
$\gamma,\Gamma_+\rightarrow 0$ we get $\langle n\rangle =0$ (complete
ground state cooling as no heating channel is open) and
$V\propto\Gamma_{0}$ (the system moves down the tilted washboard
potential at the rate $\Gamma_{0}$ which conserves the number of
vibrons).

\begin{figure}
\includegraphics[width=0.45\textwidth]{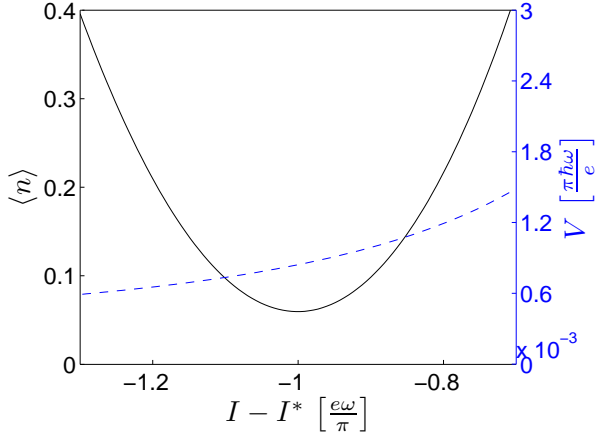}
\caption{(Color online) Average vibron population (solid) and bias
  voltage (dashed) in the stationary regime as a function of the
  current bias. Here, $\Phi=0.3$, $\Gamma=\omega/4$,
  $\mathcal{T}=\hbar\omega/20$, $n_B=20$ and $Q=$\unit[$10^5$].}
\label{final}
\end{figure}

In Fig.~\ref{final} we plot both the average stationary population of
the mechanical subsystem and the corresponding voltage drop as a
function of the bias current. As expected, the lowest occupation is
achieved when $I=I^*-e\omega/\pi$ (see Fig.~\ref{tilted}). In this
regime, we find that ground state cooling of the mechanical subsystem
is possible if the resolved side-band limit, $\omega>\Gamma$, is
achieved. Under conditions when the bias current is $I>I^*$ the
tunneling events discussed above will lead to pumping of the
mechanical subsystem, in which case the above analysis does not apply
once the limit $\mathcal{T}(\langle n\rangle+1)\sim \hbar\Gamma$ is
reached. This regime will be discussed in future work.

To conclude we have shown that a suspended nanowire which forms a weak
link in current-biased Josephson junction can be cooled to its
motional ground state. This effect derives from the coupling of the
mechanical motion of the nanowire to the electronic degrees of freedom
by a magnetic field. Furthermore, we have shown that by operating the
system under optimal bias-current conditions the occupation factor of
the vibrational modes can be greatly decreased. Also, we have found
that the potential drop over the junction might be a sensitive probe
of the stationary vibron population as it scales with the average
number of vibrons.

This work was supported in part by the Swedish VR and SSF and by the
EC project QNEMS (FP7-ICT-233952).

\appendix
\section{Derivation of density matrix}
\label{append}
The evolution of the density matrix is governed by the Liouville-von
Neumann equation \eqref{vonNeum},
\begin{align}
\label{vonNeum2}
\frac{\partial \hat{\rho}}{\partial t}=&-\frac{i}{\hbar}\left[\hat{\mathcal{H}}_0+\hat{\mathcal{H}}_{\mathcal{T}},\hat{\rho}\right]+\hat{J}(\hat{\rho})+\notag\\
&\gamma(1+n_B)\mathcal{L}_{\hat{b}}(\hat{\rho})+\gamma n_B\mathcal{L}_{\hat{b}^{\dagger}}(\hat{\rho})\,.
\end{align}
In what follows we will consider the stationary solution of
\eqref{vonNeum2} by performing a perturbative analysis in the small
parameters $\mathcal{T}/(\hbar\Gamma),\gamma/\Gamma\ll 1$. In
particular we will consider the limit of high mechanical quality
factor $Q$ such that $\gamma=\omega/Q< \mathcal{T}/(\hbar)$. To start
the analysis we take the total density matrix to be of the form,
$\hat{\rho}=\hat{\rho}_0+\epsilon\hat{\rho}_1+\epsilon^2\hat{\rho}_2...$
and equate powers of $\epsilon$. With this we find the following
equations,
\begin{equation}
\label{rho02}
0=-\frac{i}{\hbar}\left[\hat{\mathcal{H}}_0,\hat{\rho}_0\right]+\hat{J}(\hat{\rho}_0)\,,\qquad\qquad\qquad\,\,\,\,\,\,\,\,\, O(\epsilon^0)\\
\end{equation}
\begin{equation}
\label{rho12}
0=-\frac{i}{\hbar}\left[\hat{\mathcal{H}}_0,\hat{\rho}_1\right]+\hat{J}(\hat{\rho}_1)-\frac{i}{\hbar}\left[\hat{\mathcal{H}}_{\mathcal{T}},\hat{\rho}_0\right]\,,\,\,\,\,\,\,\, O(\epsilon^1)
\end{equation}
\begin{align}
0=&-\frac{i}{\hbar}\left[\hat{\mathcal{H}}_0,\hat{\rho}_2\right]+\hat{J}(\hat{\rho}_2)-\frac{i}{\hbar}\left[\hat{\mathcal{H}}_{\mathcal{T}},\hat{\rho}_1\right]+&\notag\\
\label{rho22}
&\gamma(1+n_B)\mathcal{L}_{\hat{b}}(\hat{\rho}_0)+\gamma n_B\mathcal{L}_{\hat{b}^{\dagger}}(\hat{\rho}_0)\,.\, &O(\epsilon^2)
\end{align}
Solving the above equations at each order of $\epsilon$ we find
$\hat{\rho}_0=\sum_{l,n}\vert
l,\downarrow,n\rangle\rho_0(l,\downarrow,n)\langle
l,\downarrow,n\vert$ which satisfies \eqref{rho02}. Similarly, the
first order correction to the stationary density matrix is determined
from \eqref{rho12} as,
\begin{gather}
\hat{\rho}_1=\sum_{\substack{l,n\\j=-1,0,1}}\vert l+1,\uparrow,n+i\rangle c_j(l,n)\langle l,\downarrow,n\vert+\textrm{h.c.}\,,\notag\\
\label{cis}
c_j(l,n)=\frac{\mathcal{T}_j^{(n)}\rho_0(l,\downarrow,n)}{-\Delta\mathcal{F}_j+i\hbar\Gamma/2}\,,
\end{gather}
Substituting this into \eqref{rho22} we find the equation for the
coefficients $\rho_0$ by tracing out the spin ($\uparrow,\downarrow$)
degrees of freedom,
\begin{align}
\Gamma\sum_{j=-1,0,1}\frac{4}{4(\Delta\mathcal{F}_j)^2+\hbar^2\Gamma^2}&\bigg(\vert\mathcal{T}_j^{(n-j)}\vert^2\rho_0(l-1,\downarrow,n-j)-\notag\\
&\qquad\vert \mathcal{T}_j^{(n)}\vert^2\rho_0(l,\downarrow,n)\bigg)=\notag\\
\gamma(1+n_B)[n\rho_0(l,\downarrow,n)&-(n+1)\rho_0(l,\downarrow,n+1)]+\notag\\
\label{rho0s}
\gamma n_B[(n+1)\rho_0(l,\downarrow,n)&-n\rho_0(l,\downarrow,n-1)]\,.
\end{align}
Tracing out the valley index $l$ we recover the expressions
presented in the paper, i.e. equation \eqref{cis} gives
\begin{gather}
C_j(n)\equiv \sum_{l=-\infty}^{\infty}c_j(l,n)=\frac{\mathcal{T}_j^{(n)}P(n)}{-\Delta\mathcal{F}_j+i\hbar\Gamma/2}\,,\notag
\end{gather}
whereas equation \eqref{rho0s} gives,
\begin{align}
\label{Pn}
(\Gamma_-+&\gamma(1+n_B))[(n+1)P(n+1)-nP(n)]+\notag\\
&(\Gamma_++\gamma n_B)[nP(n-1)-(n+1)P(n)]=0\,.
\end{align}
In this expression the relationship between the coefficients are,
\begin{gather}
\Gamma_{\pm}=\Gamma\frac{4\Phi^2\mathcal{T}^2}{4(\Delta\mathcal{F}_\pm)^2+\hbar^2\Gamma^2}\,,\qquad \Gamma_{0}=\Gamma\frac{4\mathcal{T}^2}{4(\Delta\mathcal{F}_0)^2+\hbar^2\Gamma^2}\,,\notag
\end{gather}
\begin{gather}
\frac{i}{\hbar}\mathcal{T}_j^{(n)}\left(C_j(n)-C_j^*(n)\right)=P(n)\Gamma_jN\,,\notag
\end{gather}
\begin{gather}
N=\left\{\begin{array}{lcl} n+1 &  & j=+\\ 1 & & j=0 \\ n & & j=-\,.\end{array}\right.\notag
\end{gather}
In the above we note that \eqref{Pn} gives the balanced equation for
the probability $P(n)$ of finding the oscillating nanowire in the
state $n$. The stationary average distribution of the vibrational
modes is then given by the solution to this equation,
\begin{gather}
\langle n\rangle=\sum_{n=0}^{\infty}nP(n) =\frac{n_B\gamma+\Gamma_+}{\Gamma_-+\gamma-\Gamma_+}\,.\notag
\end{gather}
The density matrix $\hat{\rho}_1$ allow us to evaluate the potential
drop over the junction in the stationary regime. Following the
derivation outlined in the paper we find that the lowest order term of
the density matrix, $\hat{\rho}_0$, does not contribute to the
potential drop as it is diagonal in the spin basis. As such, the
potential drop is uniquely determined from $\hat{\rho}_1$.

\bibliography{Paper}

\begin{thebibliography}{16}%
\makeatletter
\providecommand \@ifxundefined [1]{%
 \@ifx{#1\undefined}
}%
\providecommand \@ifnum [1]{%
 \ifnum #1\expandafter \@firstoftwo
 \else \expandafter \@secondoftwo
 \fi
}%
\providecommand \@ifx [1]{%
 \ifx #1\expandafter \@firstoftwo
 \else \expandafter \@secondoftwo
 \fi
}%
\providecommand \natexlab [1]{#1}%
\providecommand \enquote  [1]{``#1''}%
\providecommand \bibnamefont  [1]{#1}%
\providecommand \bibfnamefont [1]{#1}%
\providecommand \citenamefont [1]{#1}%
\providecommand \href@noop [0]{\@secondoftwo}%
\providecommand \href [0]{\begingroup \@sanitize@url \@href}%
\providecommand \@href[1]{\@@startlink{#1}\@@href}%
\providecommand \@@href[1]{\endgroup#1\@@endlink}%
\providecommand \@sanitize@url [0]{\catcode `\\12\catcode `\$12\catcode
  `\&12\catcode `\#12\catcode `\^12\catcode `\_12\catcode `\%12\relax}%
\providecommand \@@startlink[1]{}%
\providecommand \@@endlink[0]{}%
\providecommand \url  [0]{\begingroup\@sanitize@url \@url }%
\providecommand \@url [1]{\endgroup\@href {#1}{\urlprefix }}%
\providecommand \urlprefix  [0]{URL }%
\providecommand \Eprint [0]{\href }%
\@ifxundefined \urlstyle {%
  \providecommand \doi  [0]{\begingroup \@sanitize@url \@doi}%
  \providecommand \@doi [1]{\endgroup \@@startlink {\doibase
  #1}doi:\discretionary {}{}{}#1\@@endlink }%
}{%
  \providecommand \doi  [0]{doi:\discretionary{}{}{}\begingroup
  \urlstyle{rm}\Url }%
}%
\providecommand \doibase [0]{http://dx.doi.org/}%
\providecommand \Doi [0]{\begingroup \@sanitize@url \@Doi }%
\providecommand \@Doi  [1]{\endgroup\@@startlink{\doibase#1}\@@Doi}%
\providecommand \@@Doi [1]{#1\@@endlink}%
\providecommand \selectlanguage [0]{\@gobble}%
\providecommand \bibinfo  [0]{\@secondoftwo}%
\providecommand \bibfield  [0]{\@secondoftwo}%
\providecommand \translation [1]{[#1]}%
\providecommand \BibitemOpen [0]{}%
\providecommand \bibitemStop [0]{}%
\providecommand \bibitemNoStop [0]{.\EOS\space}%
\providecommand \EOS [0]{\spacefactor3000\relax}%
\providecommand \BibitemShut  [1]{\csname bibitem#1\endcsname}%
\bibitem [{\citenamefont {Schwab}\ and\ \citenamefont
  {Roukes}(2005)}]{Schwab2005}%
  \BibitemOpen
  \bibfield  {author} {\bibinfo {author} {\bibfnamefont {K.~C.}\ \bibnamefont
  {Schwab}}\ and\ \bibinfo {author} {\bibfnamefont {M.~L.}\ \bibnamefont
  {Roukes}},\ }\href@noop {} {\bibfield  {journal} {\bibinfo  {journal} {Phys.
  Today},\ }\textbf {\bibinfo {volume} {58}},\ \bibinfo {pages} {36} (\bibinfo
  {year} {2005})}\BibitemShut {NoStop}%
\bibitem [{\citenamefont {Blencowe}(2005)}]{Blencowenano}%
  \BibitemOpen
  \bibfield  {author} {\bibinfo {author} {\bibfnamefont {M.~P.}\ \bibnamefont
  {Blencowe}},\ }\Doi {10.1080/00107510500146865} {\bibfield  {journal}
  {\bibinfo  {journal} {Contemp. Phys.},\ }\textbf {\bibinfo {volume} {46}},\
  \bibinfo {pages} {249} (\bibinfo {year} {2005})}\BibitemShut {NoStop}%
\bibitem [{\citenamefont {Blencowe}(2004)}]{Blencowe2004}%
  \BibitemOpen
  \bibfield  {author} {\bibinfo {author} {\bibfnamefont {M.}~\bibnamefont
  {Blencowe}},\ }\Doi {10.1016/j.physrep.2003.12.005} {\bibfield  {journal}
  {\bibinfo  {journal} {Phys. Rep.},\ }\textbf {\bibinfo {volume} {395}},\
  \bibinfo {pages} {159} (\bibinfo {year} {2004})}\BibitemShut {NoStop}%
\bibitem [{\citenamefont {O'Connell}\ \emph {et~al.}(2010)\citenamefont
  {O'Connell}, \citenamefont {Hofheinz}, \citenamefont {Ansmann}, \citenamefont
  {Bialczak}, \citenamefont {Lenander}, \citenamefont {Lucero}, \citenamefont
  {Neeley}, \citenamefont {Sank}, \citenamefont {Wang}, \citenamefont {Weides},
  \citenamefont {Wenner}, \citenamefont {Martinis},\ and\ \citenamefont
  {Cleland}}]{OConnell2010}%
  \BibitemOpen
  \bibfield  {author} {\bibinfo {author} {\bibfnamefont {A.~D.}\ \bibnamefont
  {O'Connell}}, \bibinfo {author} {\bibfnamefont {M.}~\bibnamefont {Hofheinz}},
  \bibinfo {author} {\bibfnamefont {M.}~\bibnamefont {Ansmann}}, \bibinfo
  {author} {\bibfnamefont {R.~C.}\ \bibnamefont {Bialczak}}, \bibinfo {author}
  {\bibfnamefont {M.}~\bibnamefont {Lenander}}, \bibinfo {author}
  {\bibfnamefont {E.}~\bibnamefont {Lucero}}, \bibinfo {author} {\bibfnamefont
  {M.}~\bibnamefont {Neeley}}, \bibinfo {author} {\bibfnamefont
  {D.}~\bibnamefont {Sank}}, \bibinfo {author} {\bibfnamefont {H.}~\bibnamefont
  {Wang}}, \bibinfo {author} {\bibfnamefont {M.}~\bibnamefont {Weides}},
  \bibinfo {author} {\bibfnamefont {J.}~\bibnamefont {Wenner}}, \bibinfo
  {author} {\bibfnamefont {J.~M.}\ \bibnamefont {Martinis}}, \ and\ \bibinfo
  {author} {\bibfnamefont {A.~N.}\ \bibnamefont {Cleland}},\ }\Doi
  {10.1038/nature08967} {\bibfield  {journal} {\bibinfo  {journal} {Nature},\
  }\textbf {\bibinfo {volume} {464}},\ \bibinfo {pages} {697} (\bibinfo {year}
  {2010})}\BibitemShut {NoStop}%
\bibitem [{\citenamefont {Martin}\ \emph {et~al.}(2004)\citenamefont {Martin},
  \citenamefont {Shnirman}, \citenamefont {Tian},\ and\ \citenamefont
  {Zoller}}]{Martin2004}%
  \BibitemOpen
  \bibfield  {author} {\bibinfo {author} {\bibfnamefont {I.}~\bibnamefont
  {Martin}}, \bibinfo {author} {\bibfnamefont {A.}~\bibnamefont {Shnirman}},
  \bibinfo {author} {\bibfnamefont {L.}~\bibnamefont {Tian}}, \ and\ \bibinfo
  {author} {\bibfnamefont {P.}~\bibnamefont {Zoller}},\ }\Doi
  {10.1103/PhysRevB.69.125339} {\bibfield  {journal} {\bibinfo  {journal}
  {Phys. Rev. B},\ }\textbf {\bibinfo {volume} {69}},\ \bibinfo {pages}
  {125339} (\bibinfo {year} {2004})}\BibitemShut {NoStop}%
\bibitem [{\citenamefont {Ouyang}\ \emph {et~al.}(2009)\citenamefont {Ouyang},
  \citenamefont {You},\ and\ \citenamefont {Nori}}]{Ouyang2009}%
  \BibitemOpen
  \bibfield  {author} {\bibinfo {author} {\bibfnamefont {S.~H.}\ \bibnamefont
  {Ouyang}}, \bibinfo {author} {\bibfnamefont {J.~Q.}\ \bibnamefont {You}}, \
  and\ \bibinfo {author} {\bibfnamefont {F.}~\bibnamefont {Nori}},\ }\Doi
  {10.1103/PhysRevB.79.075304} {\bibfield  {journal} {\bibinfo  {journal}
  {Phys. Rev. B},\ }\textbf {\bibinfo {volume} {79}},\ \bibinfo {pages}
  {075304} (\bibinfo {year} {2009})}\BibitemShut {NoStop}%
\bibitem [{\citenamefont {Wilson-Rae}\ \emph {et~al.}(2004)\citenamefont
  {Wilson-Rae}, \citenamefont {Zoller},\ and\ \citenamefont
  {Imamo\={g}lu}}]{WilsonRae2004}%
  \BibitemOpen
  \bibfield  {author} {\bibinfo {author} {\bibfnamefont {I.}~\bibnamefont
  {Wilson-Rae}}, \bibinfo {author} {\bibfnamefont {P.}~\bibnamefont {Zoller}},
  \ and\ \bibinfo {author} {\bibfnamefont {A.}~\bibnamefont {Imamo\={g}lu}},\
  }\Doi {10.1103/PhysRevLett.92.075507} {\bibfield  {journal} {\bibinfo
  {journal} {Phys. Rev. Lett.},\ }\textbf {\bibinfo {volume} {92}},\ \bibinfo
  {pages} {075507} (\bibinfo {year} {2004})}\BibitemShut {NoStop}%
\bibitem [{\citenamefont {Zippilli}\ \emph {et~al.}(2009)\citenamefont
  {Zippilli}, \citenamefont {Morigi},\ and\ \citenamefont
  {Bachtold}}]{Zippilli2009}%
  \BibitemOpen
  \bibfield  {author} {\bibinfo {author} {\bibfnamefont {S.}~\bibnamefont
  {Zippilli}}, \bibinfo {author} {\bibfnamefont {G.}~\bibnamefont {Morigi}}, \
  and\ \bibinfo {author} {\bibfnamefont {A.}~\bibnamefont {Bachtold}},\ }\Doi
  {10.1103/PhysRevLett.102.096804} {\bibfield  {journal} {\bibinfo  {journal}
  {Phys. Rev. Lett.},\ }\textbf {\bibinfo {volume} {102}},\ \bibinfo {pages}
  {096804} (\bibinfo {year} {2009})}\BibitemShut {NoStop}%
\bibitem [{\citenamefont {Sonne}\ \emph {et~al.}(2010)\citenamefont {Sonne},
  \citenamefont {{M. E. Pe\~{n}a-Aza}}, \citenamefont {Gorelik}, \citenamefont
  {Shekhter},\ and\ \citenamefont {Jonson}}]{Sonne2010}%
  \BibitemOpen
  \bibfield  {author} {\bibinfo {author} {\bibfnamefont {G.}~\bibnamefont
  {Sonne}}, \bibinfo {author} {\bibnamefont {{M. E. Pe\~{n}a-Aza}}}, \bibinfo
  {author} {\bibfnamefont {L.~Y.}\ \bibnamefont {Gorelik}}, \bibinfo {author}
  {\bibfnamefont {R.~I.}\ \bibnamefont {Shekhter}}, \ and\ \bibinfo {author}
  {\bibfnamefont {M.}~\bibnamefont {Jonson}},\ }\Doi
  {10.1103/PhysRevLett.104.226802} {\bibfield  {journal} {\bibinfo  {journal}
  {Phys. Rev. Lett.},\ }\textbf {\bibinfo {volume} {104}},\ \bibinfo {pages}
  {226802} (\bibinfo {year} {2010})}\BibitemShut {NoStop}%
\bibitem [{\citenamefont {Schliesser}\ \emph {et~al.}(2008)\citenamefont
  {Schliesser}, \citenamefont {Riviere}, \citenamefont {Anetsberger},
  \citenamefont {Arcizet},\ and\ \citenamefont {Kippenberg}}]{Schliesser2008}%
  \BibitemOpen
  \bibfield  {author} {\bibinfo {author} {\bibfnamefont {A.}~\bibnamefont
  {Schliesser}}, \bibinfo {author} {\bibfnamefont {R.}~\bibnamefont {Riviere}},
  \bibinfo {author} {\bibfnamefont {G.}~\bibnamefont {Anetsberger}}, \bibinfo
  {author} {\bibfnamefont {O.}~\bibnamefont {Arcizet}}, \ and\ \bibinfo
  {author} {\bibfnamefont {T.~J.}\ \bibnamefont {Kippenberg}},\ }\Doi
  {10.1038/nphys939} {\bibfield  {journal} {\bibinfo  {journal} {Nat. Phys.},\
  }\textbf {\bibinfo {volume} {4}},\ \bibinfo {pages} {415} (\bibinfo {year}
  {2008})}\BibitemShut {NoStop}%
\bibitem [{\citenamefont {Sonne}\ \emph {et~al.}(2008)\citenamefont {Sonne},
  \citenamefont {Shekhter}, \citenamefont {Gorelik}, \citenamefont {Kulinich},\
  and\ \citenamefont {Jonson}}]{drivenoscillator}%
  \BibitemOpen
  \bibfield  {author} {\bibinfo {author} {\bibfnamefont {G.}~\bibnamefont
  {Sonne}}, \bibinfo {author} {\bibfnamefont {R.~I.}\ \bibnamefont {Shekhter}},
  \bibinfo {author} {\bibfnamefont {L.~Y.}\ \bibnamefont {Gorelik}}, \bibinfo
  {author} {\bibfnamefont {S.~I.}\ \bibnamefont {Kulinich}}, \ and\ \bibinfo
  {author} {\bibfnamefont {M.}~\bibnamefont {Jonson}},\ }\Doi
  {10.1103/PhysRevB.78.144501} {\bibfield  {journal} {\bibinfo  {journal}
  {Phys. Rev. B},\ }\textbf {\bibinfo {volume} {78}},\ \bibinfo {pages}
  {144501} (\bibinfo {year} {2008})}\BibitemShut {NoStop}%
\bibitem [{\citenamefont {Ingold}\ and\ \citenamefont
  {Nazarov}(1992)}]{IngoldNazarov}%
  \BibitemOpen
  \bibfield  {author} {\bibinfo {author} {\bibfnamefont {G.-L.}\ \bibnamefont
  {Ingold}}\ and\ \bibinfo {author} {\bibfnamefont {Y.~V.}\ \bibnamefont
  {Nazarov}},\ }\enquote {\bibinfo {title} {Single charge tunneling},}\ \
  (\bibinfo  {publisher} {Plenum Press, New York},\ \bibinfo {year} {1992})\
  p.~\bibinfo {pages} {21}\BibitemShut {NoStop}%
\bibitem [{\citenamefont {Shekhter}\ \emph {et~al.}(2006)\citenamefont
  {Shekhter}, \citenamefont {Gorelik}, \citenamefont {Glazman},\ and\
  \citenamefont {Jonson}}]{Shekhter}%
  \BibitemOpen
  \bibfield  {author} {\bibinfo {author} {\bibfnamefont {R.~I.}\ \bibnamefont
  {Shekhter}}, \bibinfo {author} {\bibfnamefont {L.~Y.}\ \bibnamefont
  {Gorelik}}, \bibinfo {author} {\bibfnamefont {L.~I.}\ \bibnamefont
  {Glazman}}, \ and\ \bibinfo {author} {\bibfnamefont {M.}~\bibnamefont
  {Jonson}},\ }\Doi {10.1103/PhysRevLett.97.156801} {\bibfield  {journal}
  {\bibinfo  {journal} {Phys. Rev. Lett.},\ }\textbf {\bibinfo {volume} {97}},\
  \bibinfo {pages} {156801} (\bibinfo {year} {2006})}\BibitemShut {NoStop}%
\bibitem [{\citenamefont {Schmidt}\ \emph {et~al.}(1991)\citenamefont
  {Schmidt}, \citenamefont {Cleland},\ and\ \citenamefont
  {Clarke}}]{Schmidt1991}%
  \BibitemOpen
  \bibfield  {author} {\bibinfo {author} {\bibfnamefont {J.~M.}\ \bibnamefont
  {Schmidt}}, \bibinfo {author} {\bibfnamefont {A.~N.}\ \bibnamefont
  {Cleland}}, \ and\ \bibinfo {author} {\bibfnamefont {J.}~\bibnamefont
  {Clarke}},\ }\Doi {10.1103/PhysRevB.43.229} {\bibfield  {journal} {\bibinfo
  {journal} {Phys. Rev. B},\ }\textbf {\bibinfo {volume} {43}},\ \bibinfo
  {pages} {229} (\bibinfo {year} {1991})}\BibitemShut {NoStop}%
\bibitem [{\citenamefont {Esteve}\ \emph {et~al.}(1986)\citenamefont {Esteve},
  \citenamefont {Devoret},\ and\ \citenamefont {Martinis}}]{Esteve1986}%
  \BibitemOpen
  \bibfield  {author} {\bibinfo {author} {\bibfnamefont {D.}~\bibnamefont
  {Esteve}}, \bibinfo {author} {\bibfnamefont {M.~H.}\ \bibnamefont {Devoret}},
  \ and\ \bibinfo {author} {\bibfnamefont {J.~M.}\ \bibnamefont {Martinis}},\
  }\Doi {10.1103/PhysRevB.34.158} {\bibfield  {journal} {\bibinfo  {journal}
  {Phys. Rev. B},\ }\textbf {\bibinfo {volume} {34}},\ \bibinfo {pages} {158}
  (\bibinfo {year} {1986})}\BibitemShut {NoStop}%
\bibitem [{\citenamefont {Hatakenaka}\ \emph {et~al.}(1990)\citenamefont
  {Hatakenaka}, \citenamefont {Takayanagi},\ and\ \citenamefont
  {Kurihara}}]{Hatakenaka1990}%
  \BibitemOpen
  \bibfield  {author} {\bibinfo {author} {\bibfnamefont {N.}~\bibnamefont
  {Hatakenaka}}, \bibinfo {author} {\bibfnamefont {H.}~\bibnamefont
  {Takayanagi}}, \ and\ \bibinfo {author} {\bibfnamefont {S.}~\bibnamefont
  {Kurihara}},\ }\Doi {10.1016/S0921-4526(09)80051-3} {\bibfield  {journal}
  {\bibinfo  {journal} {Physica B},\ }\textbf {\bibinfo {volume} {165-166}},\
  \bibinfo {pages} {931} (\bibinfo {year} {1990})}\BibitemShut {NoStop}%
\end{thebibliography}%
\end{document}